# Alfvén waves: To the 80th anniversary of discovery

A.V. Guglielmi[1], B.I. Klain[2], A.S. Potapov[3]
[1]*Schmidt Institute of Physics of the Earth RAS, Moscow, Russia, guglielmi@mail.ru*
[2]*Borok Geophysical Observatory the Branch of Schmidt Institute of Physics of the Earth RAS, Borok, Russia, klb314@mail.ru*
[3]*Institute of Solar-Terrestrial Physics SB RAS, Irkutsk, Russia, potapov@iszf.irk.ru*

**Abstract.** The paper is dedicated to the anniversary of the discovery of Alfvén waves. The concept of Alfvén waves has played an outstanding role in the formation and development of cosmical electrodynamics. A distinctive feature of Alfvén waves is that at each point in space the group velocity vector and the external magnetic field vector are collinear to each other. As a result, Alfvén waves can carry momentum, energy, and information over long distances. We briefly describe two Alfvén resonators, one of which is formed in the ionosphere, and the second presumably exists in the Earth's radiation belt. The existence of an ionospheric resonator is justified theoretically and confirmed by numerous observations. The second resonator is located between reflection points located high above the Earth symmetrically with respect to the plane of the geomagnetic equator.

Waves obeying the dispersion law $\omega = c_A k_\parallel$ are called Alfvén waves after Hannes Alfvén, who discovered them 80 years ago [Alfvén, 1942]. Here, $\omega$ is the wave frequency, $k_\parallel = \mathbf{kB}/|\mathbf{B}|$, $\mathbf{k}$ is the wave vector, $\mathbf{B}$ is the external magnetic field, $c_A = B/\sqrt{4\pi\rho}$ is the Alfvén velocity, and $\rho$ is the plasma density. The concept of Alfven waves played a significant role in the formation and development of cosmical electrodynamics [Alfvén, 1952]. It is not possible in this note to point even selectively to the colossal literature devoted to Alfvén waves. In honor of the outstanding discovery, we will only briefly describe two Alfvén resonators, one of which exists in the ionosphere, and the second presumably exists in the Earth's radiation belt.

A distinctive feature of Alfvén waves is that the vector $\mathbf{B}$ and the group velocity vector $\mathbf{v} = d\omega/d\mathbf{k}$ are collinear to each other: $\mathbf{v} = \pm\mathbf{B}/\sqrt{4\pi\rho}$. The + and − signs correspond to the inequalities $\mathbf{kB} > 0$ and $\mathbf{kB} < 0$. The property of collinearity leads to surprising consequences. Alfvén waves can carry momentum, energy, and information over long distances. Recent observations indicate that quasi-monochromatic oscillations with the frequency of 3.3 MHz are carried by Alfvén waves over 150 million km, from the Sun's surface to the Earth's surface [Guglielmi, Potapov, Dovbnya, 2015; Guglielmi and Potapov, 2021].

Another consequence of collinearity is directly related to Alfvén resonators. In the approximation of geometric optics, the Alfvén wave in an inhomogeneous medium undergoes refraction, but does not lose its directivity. In other words, if, for example, $\mathbf{kB} > 0$, then this inequality can change to the opposite only if the conditions for the applicability of the geometric optics approximation are violated. Let us assume that geometric optics works in the northern hemisphere in the F2 layer of the ionosphere, and $\mathbf{kB} > 0$. The wave propagates downward and reaches a layered structure consisting of the lower layers of the ionosphere, the atmosphere, and the upper layer of the lithosphere, the thickness of which is about the skin length. Obviously, in these layers there is a sharp violation of the conditions for the applicability of geometric optics. A reflected Alfvén wave ($\mathbf{kB} < 0$) propagating upward appears. The reflected wave crosses the F2 layer and penetrates into the exosphere. Here there is a rapid increase in the Alfvén velocity due to a rapid decrease in plasma density with height. The conditions for the approximation of geometrical optics are again violated, and a downward propagating wave appears. It is easy to understand that in this way standing Alfvén waves with a harmonic structure of spectrum are formed, locked in the ionospheric resonator. Physical-mathematical calculation and observation of ULF geoelectromagnetic oscillations in the Pc1 range convincingly testify to the existence of the Alfven ionospheric resonator [Belyaev et al., 1980; Potapov et al., 2014]. It is curious that the resonator is characterized by the distribution of spectral lines over odd harmonics [Potapov, Guglielmi, Klain, 2022].

To imagine the structure of the second resonator located high above the Earth in the near-equatorial zone of the radiation belt, we will make a short digression into crystal optics [Landau, Lifshitz, 2003]. Let us consider a strongly anisotropic uniaxial crystal and reduce the permittivity tensor to a diagonal form. It follows from the Fresnel equation that the dispersion equation for an extraordinary wave is similar to the dispersion equation for an Alfvén wave [Guglielmi, 1979]. This analogy prompts the following question: how will the dispersion equation change if off-diagonal terms are taken into account in the permittivity tensor, but the tensor remains Hermitian? In other words, we want to impart gyrotropic properties to the medium and we will immediately indicate how this can be done in a magnetically active hydrogen plasma. It turns out that it is sufficient to imagine that the hydrogen plasma contains a small admixture of heavy ions, for example, oxygen ions $O^+$ [Gintsburg, 1963].

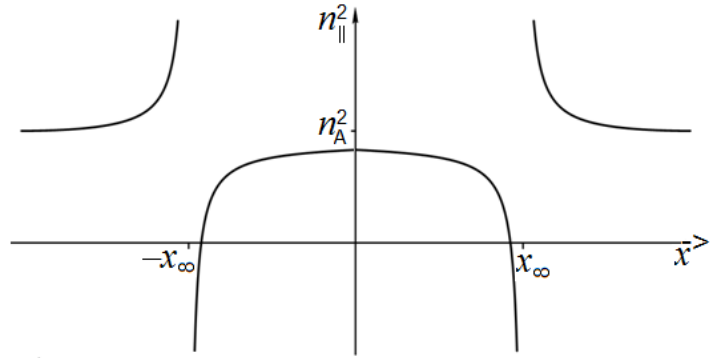

Figure. Dispersion curve in the vicinity of the equator of the geomagnetic field line (see text).

Let us use this general information to analyze Alfvén waves in the outer radiation belt in a small vicinity of the geomagnetic equator. Figure shows the dispersion curve in a hydrogen plasma containing a very small admixture of oxygen ions. The horizontal axis is directed along the geomagnetic field line. The origin of coordinates coincides with the equator of the field line. The vertical axis is the square of the longitudinal refractive index $n_\| = ck_\|/\omega$. It practically coincides with the squared refractive index of Alfvén waves $n_A = c/c_A$ everywhere, except for a narrow band in the vicinity of the resonance, at which the local gyrofrequency of the ions $O^+$ coincides with the frequency of the wave.

Let us pay attention to the fact that the turning point ($n_\| = 0$) is located somewhat to the left of the resonance. In the vicinity of turning point the geometric optics approximation is inapplicable. So a wave traveling from left to right is reflected from the turning point. The reflected wave propagates from right to left, crosses the equator and is reflected from a completely similar turning point located in the opposite hemisphere of the magnetosphere. The process is repeated and a resonator is formed [Guglielmi, Potapov, Russell, 2000] (see also [Guglielmi, Potapov, 2012, 2021; Mikhailova, 2017). Thus, the theory predicts that between the reflection points located high above the Earth symmetrically with respect to the plane of the geomagnetic equator, the equatorial Alfvén resonator is located.

The problem of excitation of the resonator deserves separate consideration. We confine ourselves here to pointing out the excitation due to the instability of the distribution of energetic particles in the outer radiation belt. The anisotropy of the velocity distribution of energetic protons leads to cyclotron instability. The nonmonotonicity of the energy distribution leads to Cherenkov instability. Let us write the Cherenkov resonance condition in the form $\omega = \mathbf{k}\mathbf{v}_D$, where $\mathbf{v}_D$ is the velocity of the azimuthal drift of energetic particles. We see that the Cherenkov excitation mechanism, generally speaking, does not depend on the type of particles. In other words, the resonator can be excited by both protons and electrons trapped in the geomagnetic trap.

In conclusion, it should be said that Alfvén's theory has radically enriched not only linear, but also nonlinear cosmic physics. In particular, satellite observations provide convincing evidence of a significant force effect of Alfvén waves on plasma in the Earth's magnetosphere [Lundin, Guglielmi, 2006].

The study was supported by the Russian Science Foundation grant No. 22-27-00280, https://rscf.ru/project/22-27-00280.